\documentclass[conference]{IEEEtran}
\IEEEoverridecommandlockouts
\usepackage{cite}
\usepackage{amsmath,amssymb,amsfonts}
\usepackage{algorithmic}
\usepackage{graphicx}
\usepackage{textcomp}
\usepackage{xcolor}
\def\BibTeX{{\rm B\kern-.05em{\sc i\kern-.025em b}\kern-.08em
		T\kern-.1667em\lower.7ex\hbox{E}\kern-.125emX}}

\usepackage{subcaption}%
\usepackage{hyperref}
\usepackage{booktabs}
\usepackage{sc24repro}

\begin{document}
	
	\title{
		Inspection of I/O Operations from System Call Traces using Directly-Follows-Graph\\
		\thanks{This research was financially supported by the J\"ulich Supercomputing Center at Forschungszentrum J\"ulich and the BMBF project 01-1H1-6013 AP6-NRW Anwenderunterst\"utzung SiVeGCS. Additionally, supervision support from RWTH Aachen University through the DFG project IRTG-2379 is greatfully acknowledged.}
	}

	\author{
		
		\IEEEauthorblockN{Aravind Sankaran\IEEEauthorrefmark{1},
			Ilya Zhukov\IEEEauthorrefmark{2}, and Wolfgang Frings\IEEEauthorrefmark{3}}
		
		\IEEEauthorblockA{\textit{J\"ulich Supercomputing Center} \\
			\textit{Forschungszentrum J\"ulich, Germany}\\
			\{\IEEEauthorrefmark{1}a.sankaran, \IEEEauthorrefmark{2}i.zhukov, \IEEEauthorrefmark{3}w.frings\}@fz-juelich.de \\
		}
		\and
		\IEEEauthorblockN{Paolo Bientinesi}
		\IEEEauthorblockA{\textit{Department of Computer Science} \\ \textit{Ume\r{a} Universitet, Sweden} \\
			pauldj@cs.umu.se}
	}

	\maketitle
	
	\begin{abstract}
		We aim to identify the differences in Input/Output (I/O) behavior between multiple user programs through the  inspection of system calls (i.e., requests made to the operating system).   
		A typical program issues a large number of I/O requests to the operating system, thereby making the process of inspection challenging. In this paper, we address this challenge by presenting a methodology to synthesize I/O system call traces into a specific type of directed graph, known as the Directly-Follows-Graph (DFG).
		Based on the DFG, we present a technique to compare the traces from multiple programs or different configurations of the same program, such that it is possible to identify the differences in the I/O behavior.
		We apply our methodology to the IOR benchmark, and compare the contentions for file accesses when the benchmark is run with different options for file output and software interface.

	\end{abstract}
	
	
	\section{Introduction}
	
	The efficiency of a computer program is often inhibited by contention for system resources. This issue is particularly evident in programs that perform significant Input/Output (I/O) operations on storage systems. In programs executed by users, requests to access system resources happen through the operating system. In this paper, we aim to analyze arbitrary user programs, without modifying them, in terms of I/O requests made to the operating system.
	
	A user program communicates with the operating system by issuing \textit{system calls}. 
	The record of the sequence of system calls made by the user program during its execution is referred to as the \textit{system trace} of that program. We analyze the contentions for system resources in programs using the system call traces. We consider the I/O-related system calls from the traces of user programs, particularly the system calls on LINUX-based operating systems that are implemented based on the interfaces defined in the C standard library (\textit{libc}) under the headers \textit{unistd.h} and \textit{sys/uio.h}.
	Performing I/O accesses directly by using the \textit{libc} calls is time-consuming and one requires an understanding of the system-specific programming requirements; for example, when porting the code to a different architecture or a Linux variant, one may have to correctly invoke the \textit{libc} calls available for that system configuration to make the I/O accesses efficient.  
	Therefore, users typically rely on standard I/O interfaces such as STDIO that manages the \textit{libc} calls under the hood\footnote{In most user programs, such as those written in FORTRAN or Python, the \textit{libc} calls are encapsulated within the software stack.}. Moreover, when it comes to parallelization of I/O accesses, users rely on more high-level interfaces (such as MPI-IO) and libraries (such as HDF5~\cite{biddiscombe_parallel_2012} and Parallel NetCDF~\cite{li_parallel_2003}). 
	The high-level interfaces are optimized for ease of use, but when to comes to achieving optimal efficiency, it has been noted that these interfaces should be used with a configuration that is tuned to a setting that is optimal for the concerned application~\cite{acharya_tuning_1996,thakur_case_1998}.
	In the process of tuning the I/O performance of a program, users must decide not only on the choice of the interface and its configuration, but also on several other parameters that should be set according to requirements of the application. 
	For example, it is important to determine the pattern of file output: whether each process should access its own file or if all processes should access a single shared file. To understand the performance impacts of the interface and configuration choices, users typically rely on I/O profiling and tracing tools to analyze their programs.
	
	Several tools exist that intercept the I/O calls from \textit{libc} to extract information and use it for analyzing and improving the I/O behavior of programs~\cite{carns_247_2009,bhatele_illuminating_2023,shende_characterizing_2012,brunst_score-p_2012, resch_vampir_2008,wang_recorder_2020,uselton_parallel_2010,niethammer_scalasca_2015}
	However, it is still challenging to perform analyses that spots \textit{differences} in I/O behavior between different configurations of a program in terms of contentions for system resources. This is mainly because each program makes a vast number of system calls through \textit{libc}, and translating a large volume of information from the system calls into a representation that facilitates precise identification of differences between the programs is not straightforward. In this paper,  we consider the problem of \textit{synthesis} of the data from system traces, i.e., combining the information in the data to extract precise insights for understanding the I/O contentions caused by a program for system resources. 
	We do not introduce yet another tool; instead, we present a methodology to synthesize the trace data into a Directly-Follows-Graph~\cite{van_der_aalst_foundations_2022} that depicts patterns of I/O system calls, which then facilitates the comparative analysis of programs in terms of requests made to the operating system. 
	The contributions of this work are the following:
		
		
	\begin{itemize}
		\item We present the theory behind the synthesis of the Directly-Follows Graph from large amounts of information in the I/O system call traces.
		
		\item We present a methodology to color the Directly-Follows-Graph, which facilitates the comparison of patterns of I/O system call accesses between several programs or multiple process running simultaneously in a program.
		
		\item We apply our methodology to the IOR benchmark, and infer the differences in file access contention when (1) several processes access a single shared file versus all the processes access their own individual files, and (2) default read and write calls in IOR are replaced with the MPI-IO counter-parts. 
	\end{itemize}

	\paragraph*{Organization}  In Sec.~\ref{sec:rel}, we review the related works. In Sec.~\ref{sec:data}, we describe the format of the trace data used as input. In Sec~\ref{sec:meth}., we present the methodology to translate the trace data into Directly-Follows Graph and explain the technique for comparative analysis. In Sec~\ref{sec:exp}, we conduct experiments and analyse the overheads, and finally in Sec.~\ref{sec:con}, we summarize the findings and draw conclusions.
	
	\section{Related Works}
	\label{sec:rel}

	In the existing tools for analyzing the I/O behavior of programs~\cite{carns_247_2009,bhatele_illuminating_2023,shende_characterizing_2012,brunst_score-p_2012, resch_vampir_2008,wang_recorder_2020,uselton_parallel_2010,niethammer_scalasca_2015}, we identify the following two common steps: (1) instrumentation of the program, which involves intercepting or interrupting calls made by the program from one or more layers of software interfaces to record relevant information, and (2) synthesis of the recorded information after the execution of the program, which involves a calculation of statistical metrics and putting them together in visualizations such as histograms, timeline plots, Gantt charts, heat maps, and more.
	
	\textbf{Instrumentation.} Most of the existing tools record information by intercepting I/O calls of the standard C library, and considerable work has been done to optimize the collection and storage of this information in formats suitable for HPC workloads.
	For example, Darshan's instrumentation is lightweight, non-intrusive and designed for 24x7 monitoring of HPC applications~\cite{carns_darshan_nodate}, while the Score-P measurement system~\cite{brunst_score-p_2012}, which collects traces in Open Trace Format Version 2 (OTF2)~\cite{otf2__developer_community_open_2023} and profiles in CUBE4~\cite{saviankou_cube_2015} formats, is tailored for scalable yet detailed program analysis. The traces and profiles generated by Score-P can be used and processed by various performance analysis tools such as Scalasca~\cite{niethammer_scalasca_2015}, TAU~\cite{shende_characterizing_2012}, Vampir~\cite{resch_vampir_2008}, and CubeGUI~\cite{saviankou_cube_2015}.
	In this work, we do not rely on the instrumentation process of any specific tool, but rather focus on introducing a method to synthesize the instrumented data. 
	To this end, we utilize the raw system call traces recorded by \textit{\textbf{strace}}~\cite{noauthor_strace_2024}—a Linux utility that uses the \textit{ptrace} system call under the hood to instrument arbitrary commands or programs in the user environment without requiring code modification. The methodology by itself does not depend on strace and can be applied over data instrumented by one of the other existing tools. 
	
	
	\textbf{Synthesis.} Extensive efforts have been made to develop different methods for synthesizing and visualizing measurement data. 
	For example, the Vampir visualization environment transforms measurements into a variety of graphical views, such as timeline plots and heat maps, with interactive elements~\cite{resch_vampir_2008}. The result of synthesis could also be a performance report that brings together information from various visualizations and statistical metrics to provide a comprehensive overview of the program. For instance, Darshan provides interfaces to synthesize their log files as static PDF reports providing an overview of the I/O performance of the program~\cite{carns_247_2009}. PyDarshan is a wrapper around Darshan that facilitates the generation of interactive HTML reports~\cite{luettgau_enabling_2023}. Drishti is a tool that synthesizes traces from DXT-Tracer to generate a variety of interactive plots~\cite{bhatele_illuminating_2023}. 
	To the best of our knowledge, we observe a lack of detailed exploration of \textit{dependency-graph-based} modeling in the  synthesis of I/O related instrumented data.
	
	Dependency-graph-based modeling has previously been used to reconstruct call-graphs, detect potential parallel regions in sequential programs (e.g., in DiscoPoP~\cite{niethammer_discopop_2015}, Parwiz~\cite{ketterlin_profiling_2012}), develop models to enable smart scheduling of HPC jobs~\cite{bienz_towards_2023}, etc. In this work, present a technique to synthesize the traces into a specific form of dependency graph known as the Directly-Follows Graph, as defined in Definition~4 in ~\cite{van_der_aalst_foundations_2022}. We discuss the computational complexities in synthesizing the graph, and then use it for comparative analysis of I/O operations across one or more programs.
	
	\section{The Input Data}
	\label{sec:data}

	We consider setups where every process involved in the execution of a program independently records the I/O system calls, and each system call captures at least the path of the accessed file, the start timestamp and the duration between start and return of the call.  In this context, processes refer to tasks executing an arbitrary command, where those tasks can be co-located on a single core, distributed across cores within a host machine, or even span multiple hosts in a parallel execution. In this section, we describe the trace records generated by strace, which are then used as inputs for our methodology.

	\subsection*{Tracing with strace}
	
\begin{figure}[t!]
		\centering
		\includegraphics[width=\linewidth]{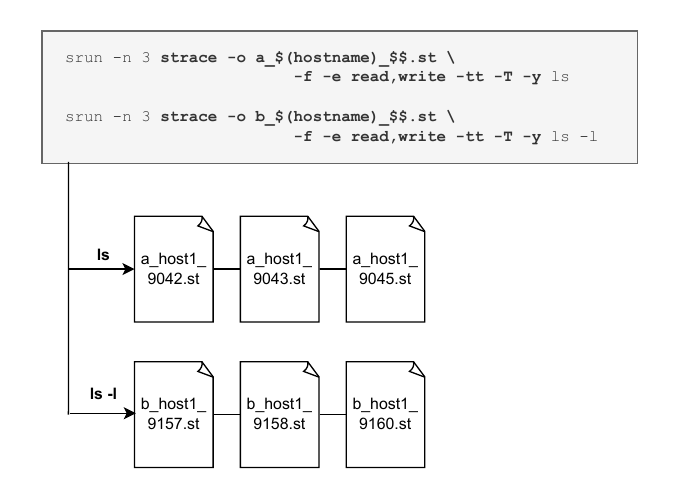}
		\caption{The commands for tracing \texttt{ls} and \texttt{ls -l} with strace, each executed on three MPI processes, generating one trace file for each process.}
		\label{fig:eg_ls}
	\end{figure}

	One could generate the traces of system calls of a command by prefixing strace to the command that needs to be traced.  In Figure~\ref{fig:eg_ls}, we show an example of running strace with \texttt{ls} and \texttt{ls -l}. Each command is run simultaneously by three MPI processes (specified as \texttt{srun -n 3}), and each MPI process records its traces in a separate file\footnote{For applications with a large number of processes, having a large number of files could lead to meta-data performance issues~\cite{alam_parallel_2011}. Therefore, after recording the traces, it is recommended that the relevant data (described in the remainder of this section) from individual trace files are parsed and combined efficiently into a suitable data format (such as a single HDF5 file). } specified by the option \texttt{- o}. 
	In order to uniquely identify each trace file, we follow a naming convention which is a combination of name of the host machine, identifier of the MPI process (\textit{rid}) and identifier of the command (\textit{cid}). In Linux, one could obtain the name of the host machine from the shell variable \texttt{hostname}. Each MPI process is represented by the identifier of the launching process, which is obtained from the variable \texttt{\$\$}.  In our example, let the commands \texttt{ls} and \texttt{ls -l} be identified with \textit{cid=}'a' and \textit{cid=}'b' respectively. 
	According to our example, the identifiers for each of the six trace files are shown in Figure~\ref{fig:eg_ls}. 
	The ASCII contents of the trace files \textit{a\_host1\_9042.st} (\textit{rid=9042} for the command \texttt{ls}) and \textit{b\_host1\_9157.st} (\textit{rid=9157} for the command \texttt{ls -l}) are shown in Figure~\ref{fig:strace_ls} and Figure~\ref{fig:strace_ls_l} respectively.

	We parse the following information from each line of each trace file:
	\begin{figure*}[h!]
		\centering
		\begin{subfigure}[b]{0.9\textwidth}
			\centering
			\includegraphics[width=1\linewidth]{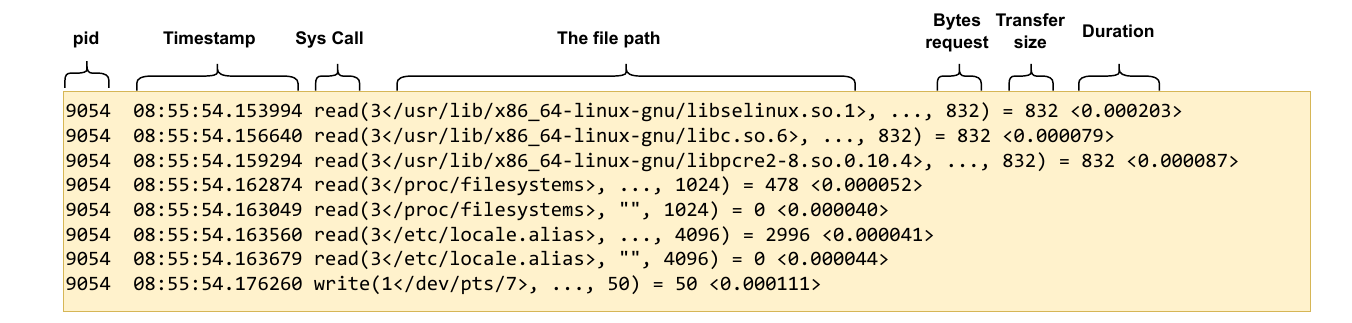}
			\caption{System calls recorded by the MPI process with the ID 9042 for the command \texttt{ls} (Trace file: \textit{a\_host1\_9042.st}).}
			\label{fig:strace_ls}
		\end{subfigure}
		\begin{subfigure}[b]{0.9\textwidth}
			\centering
			\includegraphics[width=1\linewidth]{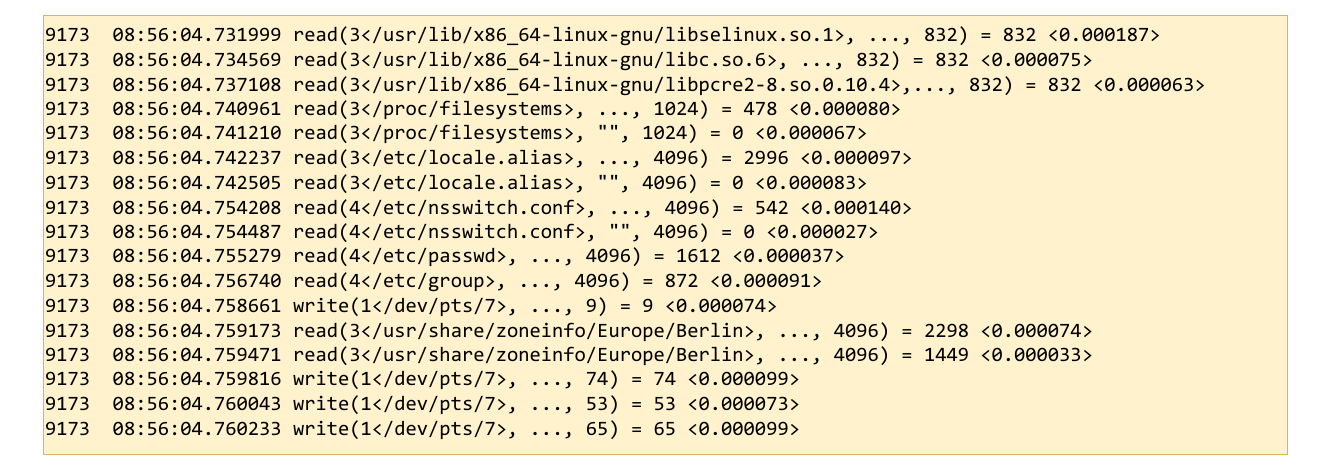}
			\caption{System calls recorded by the MPI process with the ID 9157 for the command \texttt{ls -l} (Trace file: \textit{b\_host1\_9157.st}).}
			\label{fig:strace_ls_l}
		\end{subfigure}
		\begin{subfigure}[b]{0.9\textwidth}
			\centering
			\includegraphics[width=1\linewidth]{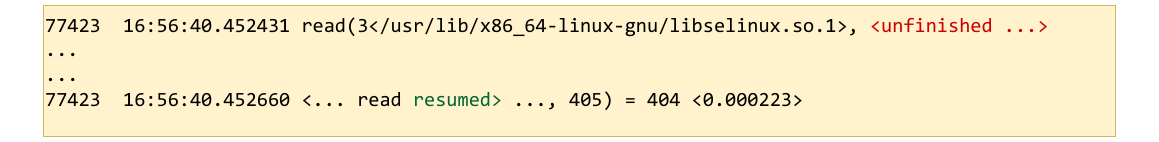}
			\caption{An example of strace records in case of simultaneous multi-processing.}
			\label{fig:strace_interrupt}
		\end{subfigure}
		
		\caption{Examples of traces generated by \texttt{strace}.}
		\label{fig:mpi_ls}
	\end{figure*}
	
		1) \textbf{The process identifier} (\textit{pid}). The identifier of the process executing the system call is recorded by specifying the option \texttt{-f}. Note that $pid$ would be different from $rid$ if the MPI process forks a child process to execute the command; in our example, $rid$ and $pid$ are different. 
		In general, for the case of Simultaneous Multi-Threading or shared memory multi-threaded applications (such as OpenMP), each MPI process could spawn more than one child processes. In the considered example, however, each MPI process is associated with only one child process.
		
		2) \textbf{The system call name} (\textit{call}).  The list of system calls to be traced is specified using the \texttt{-e} option. In our example, for simplicity, we trace only the \texttt{read} and \texttt{write} system calls. However, strace can trace additional I/O system calls, and they all can be included as input to our methodology.
		

		3) \textbf{The timestamp} (\textit{start}). The wall clock time at the start of each system call, including microseconds, is recorded by specifying the option \texttt{-tt}. For MPI processes executing in different host machines, we do \textit{not} require the system clocks to be synchronized.  
		
		4) \textbf{The duration} (\textit{dur}). The duration of a system call, which is the time between its start and return of the call, is recorded using the \texttt{-T} option. 
		
		5) \textbf{The file path} (\textit{fp}). The path of the accessed file is indicated as the first argument in the system call's signature. This is recorded using the \texttt{-y} option.
		
		6) \textbf{The transfer size} (\textit{size}). The number of bytes transferred from the page table is indicated as the return value of the call (i.e., the number after the \texttt{=} sign).  This information is parsed only for the variants of read and write system calls (and not for other I/O system calls such as \texttt{lseek}, \texttt{openat}, etc.).  Note that the number of bytes requested, which is indicated as the last argument in the system call's signature may differ from the actual number of bytes transferred.
	
	Note that for read operations on regular files, if the requested data is not in the page table, the read system call internally issues a request to obtain the data from the storage system. Read access from the storage system can be traced by recording accesses to  block-device files, which have a different directory path from regular files.
	For write operations, the system call returns as soon as the page table is updated without the guarantee of the data update being completed in the storage system; for the guarantee, one should trace the \texttt{fsync} system calls.
	
	If a system call is interrupted, it contains the  keyword \texttt{ERESTARTSYS}, and we ignore these calls. If a call is being executed and meanwhile another one is being called from a different process, then strace will preserve the order of those events and mark the ongoing call with the keyword  \texttt{<unfinished ...>}. When the call returns, it will be marked with the keyword \texttt{resumed>}. In such cases, the duration of the call and the transfer size are indicated only in the resumed record; an example of such a log in shown in Figure~\ref{fig:strace_interrupt}. The unfinished and the resumed records are matched using the $pid$, and merged into a single record. 
	
	Thus, the trace files are processed according to the requirements mentioned in this section and used as input to our methodology.
	
	\section{The DFG Synthesis}
	\label{sec:meth}

	We now explain the methodology for synthesizing the information parsed from a set of trace files into a Directly-Follows Graph (DFG). 
	Our input data can be likened to an \textit{event log} described in the field of process mining~\cite{van_der_aalst_getting_2016}.
	Formalizing the input data as an event log enables us to symbolically describe the construction of the DFG and introduce methods for comparing the I/O operations. To this end, we define the following terminologies:
	
		\textbf{Event}. Every record of system call is considered as a unique \textit{event}. An event $e$ consists of the attributes described in Sec.~\ref{sec:data}:
		\begin{equation}
			\label{eq:event}
			e = [cid, host, rid, pid, call, start, dur, fp, size ]
		\end{equation}
		The attributes $cid$, $host$ and $rid$ are inferred from the name of the trace file, and the other attributes are parsed from the records of the corresponding trace file. Let $\mathcal{E} = \{e_1, \dots, e_m\}$ be the set of all possible events in all the trace files under consideration. Then, $\nexists e_1, e_2 \in \mathcal{E}$ such that $e_1 = e_2$; i.e., no two events are exactly the same. For example, if there are two events from the same command that indicate read access of same size to the same file at the same time for the same duration from the same MPI process from the same host, then these two events \textit{must} have different $pid$. For instance, if the {\tt -f} option is not added to {\tt strace}, the \textit{pid} is not recorded. This can result in two independent invocations of system calls being identical and pointing to the same event, which is not desired.
		
		\textbf{Case}. A group of events that belong to a particular $rid$, $host$ and $cid$, arranged in increasing order of their timestamps, is referred to as a \textit{case}. In other words, the group of events in each trace file is considered a unique case.
		A case $\mathbf{c}$ is indicated as an arrangement of events as follows,
		\begin{equation}
			\mathbf{c} = \langle e_1, e_2, \dots e_n \rangle
		\end{equation}
		where all $ e_i \in \mathbf{c}$ are the events that occur in the case $\mathbf{c}$, and $\forall e_i,e_{i+1} \in \mathbf{c}$, the $start$ timestamp of $e_i$ is less than or equal to that of $e_{i+1}$. 
		For example, the case corresponding to the execution of \texttt{ls} command on \textit{rid=}9042, consisting of sequence of events parsed from the trace file $a\_host1\_9042.st$, is shown in Figure~\ref{fig:strace_ls}. 
		Note that, according to this definition of case, we do not distinguish between different SMT or OpenMP processes within the same MPI process. However, one could do so by re-defining case as a group of events belonging to the same  $cid$, $host$, and $pid$ (instead of $rid$).

		\textbf{Event-log}. A set of cases $\mathcal{C} = \{\mathbf{c}_1, \dots, \mathbf{c}_n\}$ is referred to as an \textit{event-log}. For example, the following sets of cases can be considered from the experiment in Figure~\ref{fig:eg_ls}:
		\begin{equation}
			\label{eq:cases_eg}
			\begin{aligned}
				\mathcal{C}_a =& \{\textbf{a9042},\textbf{a9043},\textbf{a9045} \} \\
				\mathcal{C}_b =& \{\textbf{b9157}, \textbf{b9158}, \textbf{b9160} \} \\
				\mathcal{C}_x = & \quad \mathcal{C}_a \cup \mathcal{C}_b
			\end{aligned}
		\end{equation}
		where $\mathcal{C}_a$ is the set of cases executing the command \texttt{ls}, $\mathcal{C}_b$ is the set of cases executing the command \texttt{ls -l}, and $\mathcal{C}_x$ is the set of cases involved in the execution of both the commands. 
		
		
		\textbf{Mapping and Activity}. A \textit{mapping} is a \textit{partial} function $f$ that maps an event $e \in \mathcal{E}$ to a named entity referred to as an \textit{activity}  $a \in \mathcal{A}_f$, and it is denoted as $f: \mathcal{E} \rightharpoonup \mathcal{A}_f $.
		The mapping $f$ is a function because an event $e \in \mathcal{E}$ is mapped to at most one activity $a \in \mathcal{A}_f$, and it is partial because not all  $e \in \mathcal{E}$ are required to have a mapping. 
		For example, consider the following mapping:
		\begin{equation}
			\label{eq:mapping}
			\begin{aligned}
				\hat{f}: &\text{ for a given event, return a string concatenating} \\
				&\text{ $call$ with $fp$ truncated to contain at most  } \\
				&\text{ top two directory levels. }
			\end{aligned}
		\end{equation}
		According to this mapping, the event parsed from the first line of the trace file in Figure~\ref{fig:strace_ls_l} would map to ``\textit{\textbf{read}:/usr/lib}''. Notice that $f$ can be one-to-one or many-to-one. Therefore, the reverse mapping $f^{-1}: \mathcal{A}_f \to \mathcal{E}$ is a \textit{multi-valued} function\footnote{A general ``function'' can only be either one-to-one or many-to-one. The reverse mapping of many-to-one is one-to-many, and it is not a function in the normal sense, and therefore the term multi-valued functions is used to distinguish from normal functions.} that maps every $a \in \mathcal{A}_f$ to a subset of events in $\mathcal{E}$; i.e.,  $f^{-1}(a) \subseteq \mathcal{E}$. Following up on our example, $\hat{f}^{-1}$(\textit{\textbf{read}:/usr/lib}) is a subset of events that correspond to those in the first three lines of the trace file shown in Figure~\ref{fig:strace_ls_l}.

		\textbf{Trace}. For a given mapping $f:\mathcal{E} \rightharpoonup \mathcal{A}_f$ and a case $\mathbf{c} \in \mathcal{C}$, the sequence of activities corresponding to the events observed in $\mathbf{c}$ is called an activity trace or simply \textit{trace} $\sigma_f(\mathbf{c})$, i.e., 
		\begin{equation}
			\begin{aligned}
				\sigma_f(\mathbf{c}) &= f \circ \mathbf{c} \\
				&=  \langle f(e_1), f(e_2), \dots f(e_n) \rangle \\
				&=  \langle a_1, a_2, \dots a_n \rangle
			\end{aligned}
		\end{equation}
		For example, for the mapping $\hat{f}$ (defined in Equation~\ref{eq:mapping}), the trace for the case $\textbf{a9042} \in \mathcal{C}_a$ (corresponding to the trace file in Figure~\ref{fig:strace_ls}) is:
		\begin{equation*}
			\begin{aligned}
				\sigma_{\hat{f}}(\textbf{a9042}) = &\langle \textit{ \textbf{read}:/usr/lib}, \textit{\textbf{read}:/usr/lib}, \textit{\textbf{read}:/usr/lib},\\
				&\textit{\textbf{read}:/proc/filesystems}, \textit{\textbf{read}:/proc/filesystems}, \\
				&\textit{\textbf{read}:/etc/locale.alias}, \textit{\textbf{read}:/etc/locale.alias}, \\
				&\textit{\textbf{write}:/dev/pts} \rangle 
			\end{aligned}
		\end{equation*}
		Note that for all $ e_i, e_j \in \mathbf{c}$, $e_i$ precedes $e_j$ implies $a_i$ precedes $a_j$ for all $a_i, a_j \in \sigma_f(\mathbf{c})$.
		
		\textbf{Activity-log}. For a given event-log $\mathcal{C}$ and a mapping $f:\mathcal{E} \rightharpoonup \mathcal{A}_f$,  an \textit{activity-log} $L_{f}(\mathcal{C})$ is a multi-set (i.e., a set with multiple instances of the same element) of traces over $\mathcal{A}_f$ for $\mathcal{C}$, and it is represented as $L_f(\mathcal{C}) \in \mathbb{B}({\mathcal{A}_f}^*)$, where ${\mathcal{A}_f}^*$ is the set of all possible sequences of activities in $\mathcal{A}_f$.
		For example, consider a fictitious event-log $\mathcal{C} = \{0,1,2\}$. If $\mathcal{A}_f = \{a,b,c\}$, and the traces are $\sigma_f(0) = \langle a,a,b \rangle$, $\sigma_f(1) = \langle a,a,b \rangle$, $\sigma_f(2) = \langle a,c \rangle$, then the activity-log $L_f(\mathcal{C}) = \{\langle a,a,b \rangle^2, \langle a,c \rangle\}$; the trace $\langle a,a,b \rangle$ from  $\sigma_f(0)$ and $\sigma_f(1)$ is indicated with multiplicity 2. Now, consider the event-logs shown in Equation~\ref{eq:cases_eg} and the mapping $\hat{f}$. The activity-log $L_{\hat{f}}(\mathcal{C}_a)$ after appending every trace with a start ($\bullet$) and an end (\scalebox{0.5}{$\blacksquare$}) activity would be:
		\begin{equation*}
			\begin{aligned}
				L_{\hat{f}}(\mathcal{C}_a) = &\{ \langle \bullet,  \textit{ \textbf{read}:/usr/lib}, \textit{\textbf{read}:/usr/lib}, \textit{\textbf{read}:/usr/lib},\\
				&\textit{\textbf{read}:/proc/filesystems}, \textit{\textbf{read}:/proc/filesystems}, \\
				&\textit{\textbf{read}:/etc/locale.alias}, \textit{\textbf{read}:/etc/locale.alias}, \\
				&\textit{\textbf{write}:/dev/pts},  \text{\scalebox{0.5}{$\blacksquare$}} \rangle^3 \}
			\end{aligned}
		\end{equation*}
		From $\mathcal{C}_a$, all the three cases \textbf{a9042}, \textbf{a9043} and \textbf{a9045} map to the same trace, and hence $L_{\hat{f}}(\mathcal{C}_a)$ consist of a single trace with multiplicity of 3. Similarly, the activity logs $L_{\hat{f}}(\mathcal{C}_b)$ and $L_{\hat{f}}(\mathcal{C}_x)$ are:
		\begin{equation*}
			\begin{aligned}
				L_{\hat{f}}(\mathcal{C}_b) = &\{  \langle \bullet, \textit{\textbf{read}:/usr/lib}, \textit{\textbf{read}:/usr/lib}, \textit{\textbf{read}:/usr/lib},\\
				&\textit{\textbf{read}:/proc/filesystems}, \textit{\textbf{read}:/proc/filesystems}, \\
				&\textit{\textbf{read}:/etc/locale.alias}, \textit{\textbf{read}:/etc/locale.alias}, \\
				&\textit{\textbf{read}:/etc/nsswitch.conf}, \textit{\textbf{read}:/etc/nsswitch.conf}, \\
				&\textit{\textbf{read}:/etc/passwd}, \textit{\textbf{read}:/etc/group},\\
				&\textit{\textbf{write}:/dev/pts}, \textit{\textbf{read}:/usr/lib}, \textit{\textbf{read}:/usr/lib}, \\
				&\textit{\textbf{write}:/dev/pts}, \textit{\textbf{write}:/dev/pts}, \textit{\textbf{write}:/dev/pts},  \text{\scalebox{0.5}{$\blacksquare$}}  \rangle^3 \}
			\end{aligned}
		\end{equation*}
		\begin{equation*}
			L_{\hat{f}}(\mathcal{C}_f) = L(\mathcal{C}_a) \cup L(\mathcal{C}_b)
		\end{equation*}
	Thus, an activity-log can be seen as a query and an abstraction applied to an event-log through the mapping $f$; this mapping provides a mechanism to shift the focus of information in the event-log. The activity-log is used as an input to construct the DFG. 
	
	
	\subsection{Construction of the Directly-Follows-Graph}
	\label{sec:dfg_construction}
%
	\begin{figure*}[t!]
		\centering
		\begin{subfigure}[b]{\textwidth}
			\centering
			\includegraphics[width=0.35\linewidth]{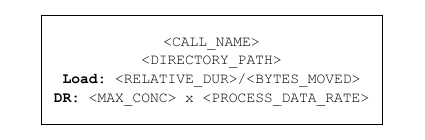}
			\caption{Semantics inside a node of the DFG.}
			\label{fig:dfg_box}
		\end{subfigure}
		\begin{subfigure}[b]{0.32\textwidth}
			\centering
			\includegraphics[width=0.65\linewidth]{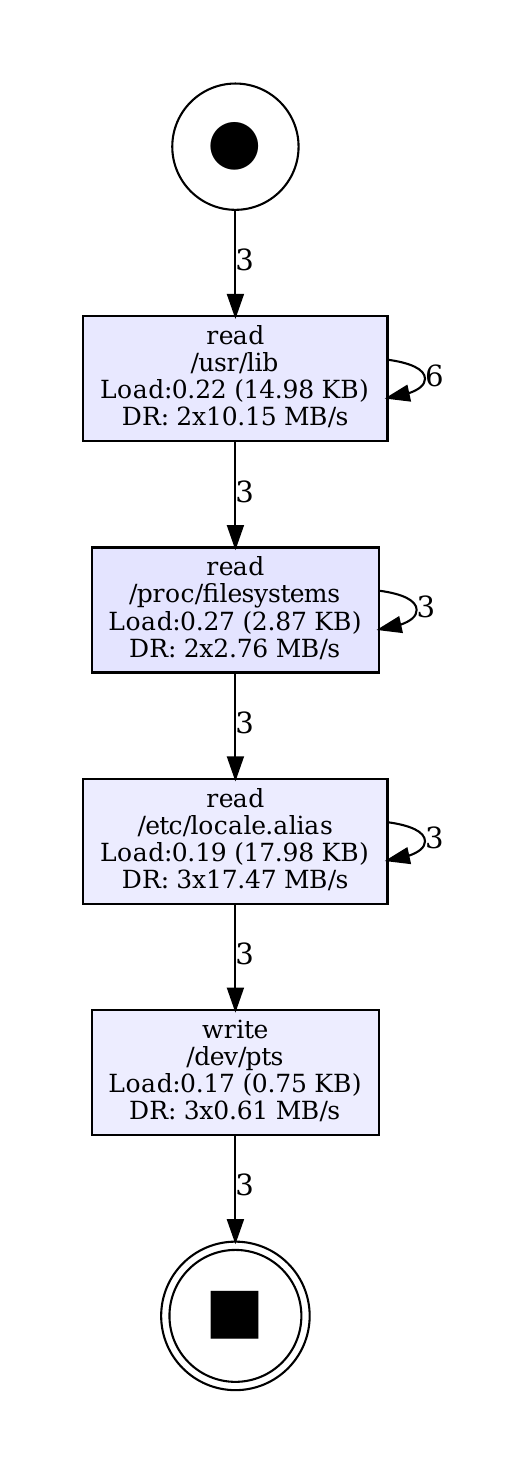}
			\caption{$G[L_{\hat{f}}(\mathcal{C}_a)]$}
			\label{fig:mpi_ls_dfg}
		\end{subfigure}
		\begin{subfigure}[b]{0.32\textwidth}
			\centering
			\includegraphics[width=0.85\linewidth]{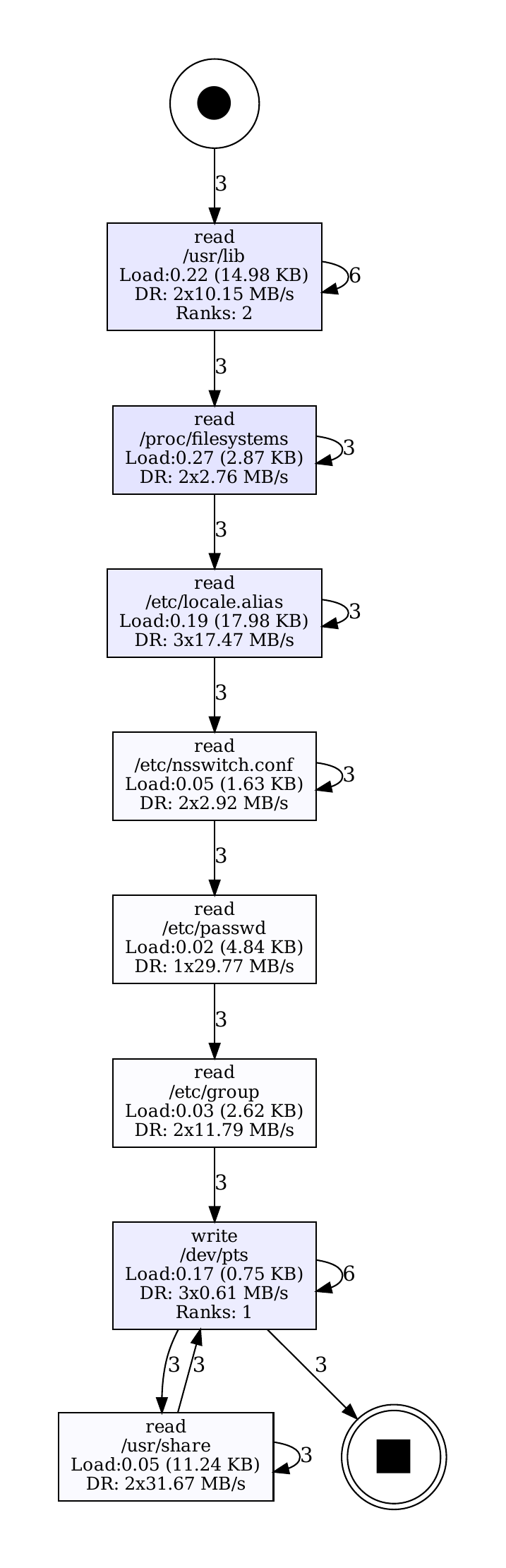}
			\caption{$G[L_{\hat{f}}(\mathcal{C}_b)]$}
			\label{fig:mpi_ls_l_dfg}
		\end{subfigure}
		\begin{subfigure}[b]{0.32\textwidth}
			\centering
			\includegraphics[width=0.9\linewidth]{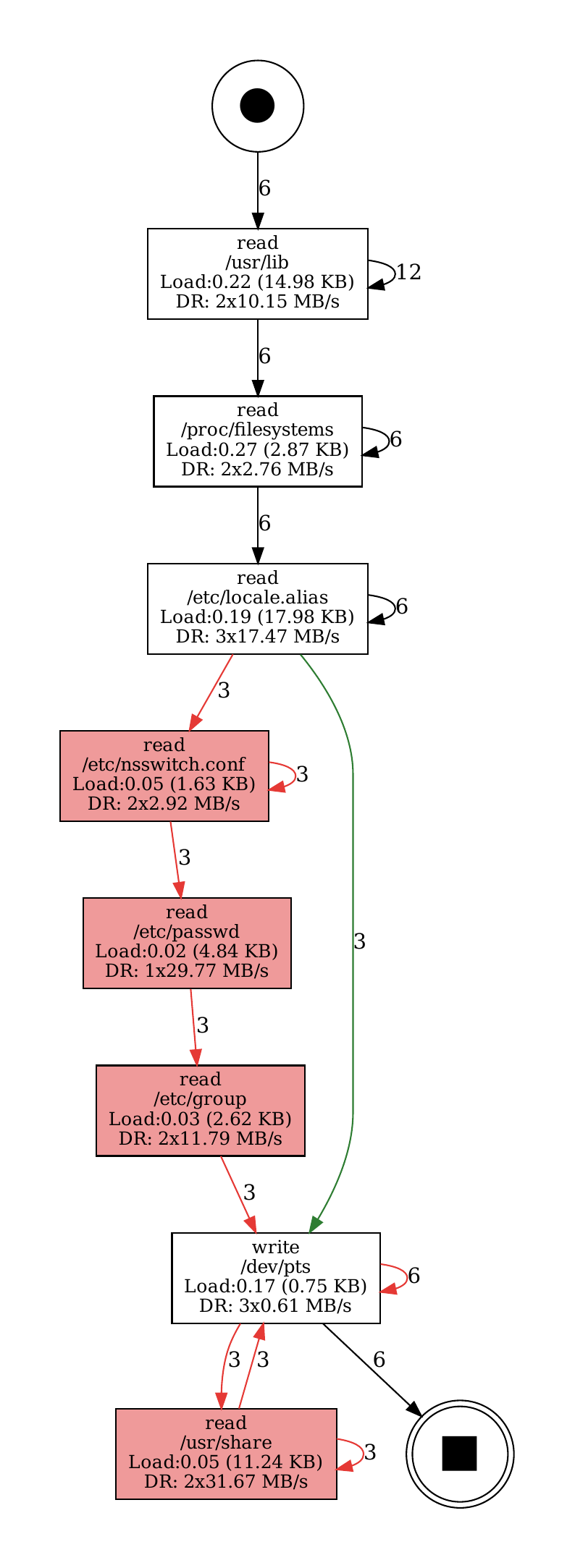}
			\caption{$G[L_{\hat{f}}(\mathcal{C}_x)]$}
			\label{fig:mpi_ls_compare_dfg}
		\end{subfigure}
		
		\caption{The DFG synthesis for the event-logs in Equation~\ref{eq:cases_eg}. The nodes indicate the file access activities and the number on the edges indicate the number of times the directly-follows relation between two activities was observed. }
		\label{fig:ls_dfg}
	\end{figure*}

	Given an event-log $\mathcal{C}$ and a mapping $f:\mathcal{E} \rightharpoonup \mathcal{A}_f$, the activity-log $L_f(\mathcal{C}) \in \mathbb{B}({\mathcal{A}_f}^*)$ is determined, and the DFG 
	$G[L_f(\mathcal{C})]$ is 
	constructed such that $a \in \mathcal{A}_f$ are the nodes and an edge $(a_1, a_2)$ where $a_1, a_2 \in \mathcal{A}_f$ exists if and only if there exists a trace in the activity-log, i.e.,  $\exists \sigma_f \in L_f$ such that $a_1$ immediately precedes (or directly follows) $a_2$. If $a_1 = a_2$, then the node has an edge pointing to itself. 
	
	For example, for the activity-logs  $L_{\hat{f}}(\mathcal{C}_a)$,  $L_{\hat{f}}(\mathcal{C}_b)$ and  $L_{\hat{f}}(\mathcal{C}_x)$, the corresponding the DFGs $G[L_{\hat{f}}(\mathcal{C}_a)]$,  $G[L_{\hat{f}}(\mathcal{C}_b)]$ and  $G[L_{\hat{f}}(\mathcal{C}_x)]$ are shown in Figure~\ref{fig:mpi_ls_dfg}, Figure~\ref{fig:mpi_ls_l_dfg} and Figure~\ref{fig:mpi_ls_compare_dfg} respectively. The number on the edges indicates how many times the corresponding directly-follows relation was observed in the activity-log. The statistics indicated in the nodes (i.e., Load and DR) and the coloring schemes will be explained in the following sub-sections. The implementation for scalable construction of $G$ from activity-log  $L_f(\mathcal{C})$  is discussed in~\cite{gaaloul_scalable_2015, evermann_scalable_2016}. Thus, $G[L_{\hat{f}}(\mathcal{C}_a)]$ is the DFG synthesis according to $\hat{f}$ for the processes executing the \texttt{ls} command, and similarly $G[L_{\hat{f}}(\mathcal{C}_b)]$ for the processes executing the \texttt{ls -l} command.  $G[L_{\hat{f}}(\mathcal{C}_x)]$ is the synthesis of events from all the processes executing both commands. 
	
\begin{figure*}[t!]
		\centering
		\includegraphics[width=0.9\linewidth]{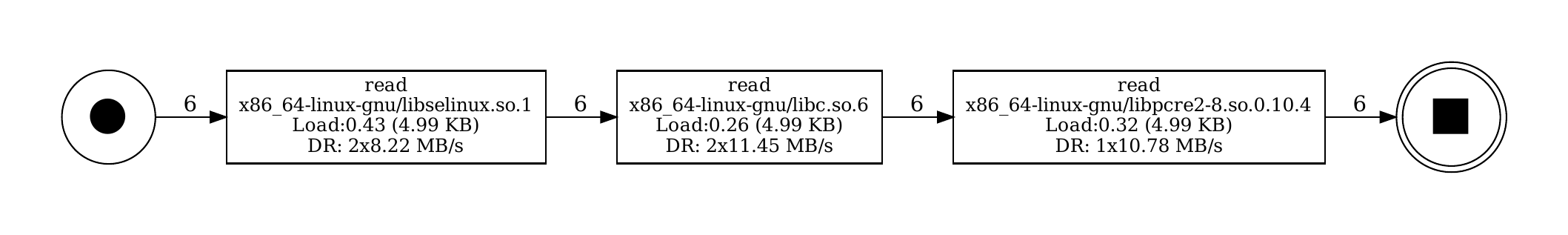}
		\caption{The DFG synthesis for the event-logs in Equation~\ref{eq:cases_eg}. The nodes indicate the file access activities and the number on the edges indicate the number of times the directly-follows relation between two activities was observed. }
		\label{fig:mpi_ls_usr_lib_dfg}
	\end{figure*}

	The DFG is a response to a query applied through $f$ on the event-log. One could modify the query to restrict the synthesis to a particular section of the event-log. For example, to restrict the synthesis to the directory \texttt{/usr/lib}, define a mapping $f_1$ such that it maps an event to an activity only if the file path contains the sub-string \textit{/usr/lib}. Then, for the corresponding activity-log $L_{f_1}(\mathcal{C}_x) \in \mathbb{B}({\mathcal{A}_{f_1}}^*)$ over $\mathcal{C}_x$ and the mapping $f_1:\mathcal{E} \rightharpoonup \mathcal{A}_{f_1}$, 
	the DFG file access footprint $G[L_{f_1}(\mathcal{C}_x)]$ is shown in Figure~\ref{fig:mpi_ls_usr_lib_dfg}.  Thus, the DFG provides a way for the users to interactively visualize the I/O accesses made by their application. 
	
	\subsection{Activity Statistics}
	\label{sec:meth_stats}

	\begin{figure}[t!]
		\centering
		\includegraphics[width=\linewidth]{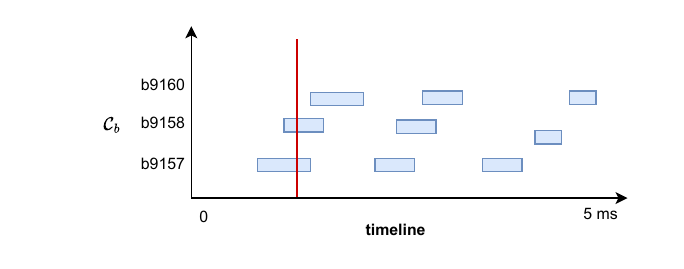}
		\caption{The timeline plot: $\mathbf{t}_{\hat{f}}(``\textit{\textbf{read}:/usr/lib}", \mathcal{C}_b)$.}
		\label{fig:timeline_uselib}
	\end{figure}

	Given an event-log $\mathcal{C}$ and a mapping $f: \mathcal{E} \rightharpoonup \mathcal{A}_f$, we compute statistics for each $a \in \mathcal{A}_f$, which add performance perspectives to the nodes of the DFG. Particularly, for each node, we aim to determine the proportion of system time spent relative to activities represented by the other nodes, the number of bytes transferred, and the rate of data movement. To this end, we compute the following statistics: 
	
	\begin{itemize}
		\setlength\itemsep{1em}
		\item \textbf{Relative duration}: The \textit{relative duration} of an activity $a \in \mathcal{A}_f$ encountered in $\mathcal{C}$ is defined as the proportion of the time spent by events on activity $a$ relative to the total time spent across all activities $\forall a \in \mathcal{A}_f$. Let $e[dur]$ be the duration of system call for event $e \in \mathcal{E}$. In order to compute the relative duration $rd_f(a, \mathcal{C})$, we first aggregate the duration of all the events related to $a \in \mathcal{A}_f$ occurring in $\mathcal{C}$; i.e., 
		\begin{equation}
			\mathbf{d}_f(a, \mathcal{C}) = \{e[dur] \ | \  \forall \mathbf{c} \in \mathcal{C}, \forall e \in \mathbf{c} \cap f^{-1}(a)\}
		\end{equation}
		In words, for a given event-log $\mathcal{C}$ and an activity $a \in \mathcal{A}_f$, we check for each case in $\mathbf{c} \in \mathcal{C}$ and for each event $e \in \mathbf{c}$, if this event $e$ is in the set of events that maps to activity $a$, i.e., $f^{-1}(a)$. If it does, then the value corresponding to the duration attribute of this event, i.e., $e[dur]$ is added to the set $\mathbf{d}_f(a, \mathcal{C})$.  
		For example,  $\mathbf{d}_{\hat{f}}(``\textit{\textbf{read}:/usr/lib}", \mathcal{C}_a)$ is a set that constitutes the duration values parsed from the three trace files represented by $\mathcal{C}_a$, and consists of only those events with file-path containing the sub-string \texttt{/usr/lib}. After determining $\mathbf{d}_f(a, \mathcal{C})$, we compute the sum of all the duration values in the set,
		\begin{equation}
			\bar{\text{d}}_f(a, \mathcal{C}) = \sum\mathbf{d}_f(a, \mathcal{C})
		\end{equation}
		and the relative duration  $rd_f(a, \mathcal{C})$) is computed as follows:
		\begin{equation}
			\label{eq:rd}
			rd_f(a, \mathcal{C})  = \frac{\bar{\text{d}}_f(a, \mathcal{C})}{\sum_{\substack{a \in \mathcal{A}_f}}\bar{\text{d}}_f(a, \mathcal{C})}
		\end{equation}
		This metric allows us to gauge the relative importance of each activity in terms of the system time spent on I/O.
		
		\item \textbf{Total bytes moved}: Let $e[size]$ be the number of bytes moved for event $e \in \mathcal{E}$. The total number of bytes moved for activity $a \in \mathcal{A}_f$ occurring in $\mathcal{C}$ is:
		\begin{equation}
			b_f(a, \mathcal{C}) = \sum\{e[size] \ | \  \forall \mathbf{c} \in \mathcal{C}, \forall e \in \mathbf{c} \cap f^{-1}(a)\}
		\end{equation}
	\end{itemize}
	
	In the nodes of the DFGs in Figure~\ref{fig:ls_dfg}, the relative duration and total bytes moves are combined and indicated as:
	\begin{equation}
		\label{eq:load}
		``\textbf{Load:} \ rd_{\hat{f}} \ (b_{\hat{f}})"
	\end{equation}
	Next, we approximate the rate of data movement for each activity by computing the following:
	
	\begin{itemize}
		\item \textbf{Process Data Rate}: The average number of bytes transferred per second per process for activity $a \in \mathcal{A}_f$ encountered in $\mathcal{C}$ is considered as the \textit{process data rate} for activity $a$, and it is denoted as $\bar{dr}_f(a, \mathcal{C})$. In order to compute the rate at which each process performing an activity moves data, we first define the data rate for an event $e \in \mathcal{E}$:
		\begin{equation}
			dr(e) = \frac{e[size]}{e[dur]}
		\end{equation}
		We then aggregate the event data rates related to activity $a$ occurring in $\mathcal{C}$:
		\begin{equation}
			\label{eq:bw_values}
			\mathbf{dr}_f(a, \mathcal{C}) = \{dr(e) \ | \  \forall \mathbf{c} \in \mathcal{C}, \forall e \in \mathbf{c} \cap f^{-1}(a)\}
		\end{equation}
		The process data rate 	$dr_f$ is then the arithmetic mean ($\mu$) of all the values in  $\mathbf{dr}_f$:
		\begin{equation}
			\bar{dr}_f(a, \mathcal{C}) = \mu(\mathbf{dr}_f(a, \mathcal{C}))
		\end{equation}
		
		\item \textbf{Max-Concurrency}: The\textit{ maximum concurrency} attained for activity $a \in \mathcal{A}_f$ encountered in $\mathcal{C}$ is the highest number of concurrent events 
		corresponding to $a$ that occurred in $\mathcal{C}$, and it is denoted as $mc_f(a, \mathcal{C})$. In order to compute $mc_f$, we first define the start and end timestamp for each event $e \in \mathcal{E}$ as a tuple:
		\begin{equation}
			t(e) = (e[start], e[start]+e[dur])
		\end{equation}
		and aggregate the time stamps of all the events for each $a$ occurring in $\mathcal{C}$ into a list:
		\begin{equation}
			\mathbf{t}_f(a, \mathcal{C}) = [t(e) \ | \  \forall \mathbf{c} \in \mathcal{C}, \forall e \in \mathbf{c} \cap f^{-1}(a)]
		\end{equation}
		Each $t(e) \in \mathbf{t}_f(a, \mathcal{C})$ can be visualized as a range from start to end timestamp in a timeline plot. For example, the timeline plot of  $\mathbf{t}_{\hat{f}}(``\textit{\textbf{read}:/usr/lib}", \mathcal{C}_b)$ is shown in Figure~\ref{fig:timeline_uselib}. The max-concurrency  $mc_f$ is computed as:
		\begin{equation}
			mc_f(a, \mathcal{C}) = \text{\texttt{get\_max\_concurrency}}(\mathbf{t}_f(a, \mathcal{C}))
		\end{equation}
		The algorithm \texttt{get\_max\_concurrency} first sorts $\mathbf{t}_f$ according to increasing start timestamps, iterates through the sorted $\mathbf{t}_f$, and determines the maximum number of consecutive events that could be identified such that the end time of the first event is greater than the start time of the last event. For example, in $\mathbf{t}_{\hat{f}}(``\textit{\textbf{read}:/usr/lib}", \mathcal{C}_b)$, max-concurrency is 2. Notice that, for precise estimation of this metric in a program with processes distributed across multiple nodes, the system clocks have to be synchronized. If they are not, then the $mc_f$ values may not be exact. However, \textit{not} having the clocks synchronized does \textit{not} affect the DFG construction or the other metrics. 
		

	\end{itemize}
	In the nodes of the DFGs in Figure~\ref{fig:ls_dfg}, the process data rate and the max-concurrency statistics are combined and indicated as:
	\begin{equation}
		``\textbf{DR: } mc_{\hat{f}} \times \bar{dr}_{\hat{f}}"
	\end{equation}
	This metric shows an estimation of the rate at which a file access activity induces I/O load on the system.
	
	Thus, appending the DFG with statistics related to file access activities enhances the visualization by providing additional information, with which one could analyse not only the file accesses but also how the activities differ from each other in terms of system load and data movements.

	
	\subsection{Performance Comparisons via Graph Coloring}
	\label{sec:meth_coloring}
	
	For a given event-log $\mathcal{C}$ and a mapping  $f: \mathcal{E} \rightharpoonup \mathcal{A}_f$, we color the nodes and edges of the DFG $G[L_f(\mathcal{C})]$ according to one of the following strategies:
	
	\begin{enumerate}
		\item \textbf{Statistics-based coloring}:  A straightforward method to visually compare the I/O operations represented by the activities in $\mathcal{A}_f$ is by coloring the nodes of the DFG $G[L_f(\mathcal{C})]$ based on the statistics described in the previous sub-section. For instance, in Figure~\ref{fig:mpi_ls_dfg} and Figure~\ref{fig:mpi_ls_l_dfg}, the activities are colored based on the relative duration; i.e., higher the value of $rd_f$, the darker the shade of blue. Alternatively, one could color the nodes based on the number of bytes $b_f$ moved. However, with this method, one could not identify the similarities and differences in the I/O operations \textit{among the different processes}, i.e., among the different cases $\mathbf{c}_i \in \mathcal{C}$. 
		
		\item \textbf{Partition-based coloring}: In order to make comparisons among the cases in an event-log $\mathcal{C}$, we perform the following steps:
		\begin{enumerate}
			\item  From the event-log $\mathcal{C}$, identify two mutually exclusive subsets $\mathcal{G}$ and $\mathcal{R}$.
			\item  Construct the DFG  $G[L_f(\mathcal{C})]$ from the full event-log $\mathcal{C}$, and the DFGs $G[L_f(\mathcal{G})]$ and $G[L_f(\mathcal{R})]$ from the event-log subsets.
			\item Color the nodes and edges of $G[L_f(\mathcal{C})]$ as follows:
			\begin{itemize}
				\item The nodes and edges that occur \textit{exclusively} in $G[L_f(\mathcal{G})]$ are given the color \textit{green}.
				\item The nodes and edges that occur \textit{exclusively} in $G[L_f(\mathcal{R})]$ are given the color \textit{red}.
				\item The nodes and edges that occur in both  $G[L_f(\mathcal{G})]$ and  $G[L_f(\mathcal{R})]$ are not colored.
			\end{itemize}
		\end{enumerate}

		For example, let us compare and contrast I/O operations between the processes executing the commands \texttt{ls} and \texttt{ls -l}. To this end, we consider the following partition of $\mathcal{C}_x$ (based on Equation~\ref{eq:cases_eg}):
		\begin{equation}
			\begin{aligned}
				\mathcal{G}_x &= \mathcal{C}_a \qquad \text{i.e., the processes executing \texttt{ls}}\\
				\mathcal{R}_x &= \mathcal{C}_a \qquad \text{i.e., the processes executing \texttt{ls -l}}
			\end{aligned}
		\end{equation}
		The coloring of the DFG $G[L_{\hat{f}}(\mathcal{C}_x)]$ based on the DFGs constructed from the subsets $\mathcal{G}_x$ and $\mathcal{R}_x$ is shown in Figure~\ref{fig:mpi_ls_compare_dfg}. 
		The nodes and edges in red are those that occur exclusively in the processes executing the command \texttt{ls -l}. There are no activities that occur exclusively in \texttt{ls}, except a single directly-follows relation indicated as an edge from ``\textit{\textbf{read}:/etc/locale.alias}'' to ``\textit{\textbf{write}:/dev/pts}''. 
		The remaining activities and relations are observed in the processes executing both commands.
	\end{enumerate}
	
	\section{Experiments}
	\label{sec:exp}
	
	In this section, we apply DFG synthesis to I/O traces from runs of IOR benchmark. First, we describe our implementation and the HPC environment on which the experiments are conducted. Then, we run the IOR benchmark with different options for file output and software interface, and compare the file access contentions between the runs.
	
\begin{figure}[t!]
		\centering
		\includegraphics[width=0.9\linewidth]{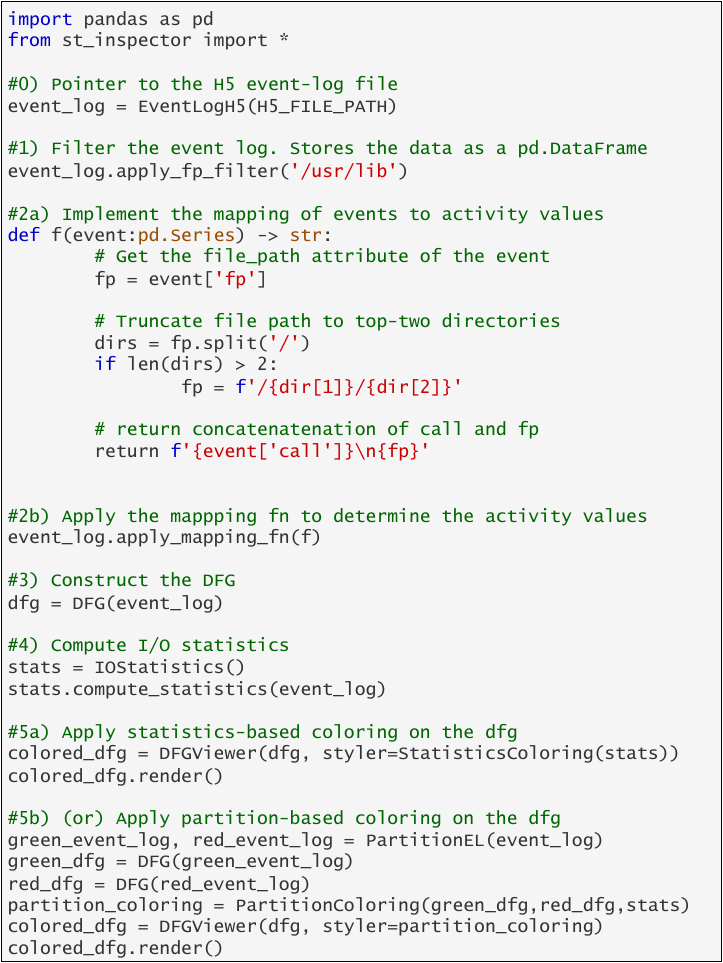}
		\caption{DFG synthesis using the Python library: \textit{st\_inspector}.}
		\label{fig:code}
	\end{figure}

	\textbf{Implementation:} We use strace version 6.4 to trace user programs. After the program execution, the individual trace files are processed as described in Sec.~\ref{sec:data} and stored in a single HDF5 file. Each processed trace file (i.e., each case) is stored in a separate group within the HDF5 file as a table. The columns of these tables correspond to the event attributes \textit{pid, call, start, dur, fp, size} as defined in Sec.~\ref{sec:data}. Each table contains the events for a particular case, sorted by increasing start timestamps (\textit{start}). This format of the HDF5 file naturally represents an event-log according to our definition in Sec.~\ref{sec:meth}. To synthesize the DFG, we perform the following steps in Python (also shown in Figure~\ref{fig:code}): 
	
	\begin{enumerate}
		\item From each table in the HDF5 file, retrieve the events containing a given sub-string (e.g., ``/usr/lib'') in the file path value (\textit{fp}). The resulting tables are concatenated and stored as a DataFrame object implemented by the Pandas library. The DataFrame additionally has a column named ``case" that points  each event to the corresponding trace file name.
		\item Implement the mapping function $f$ and apply it to the DataFrame to add a column named ``activity''. In step 2 of the code in Figure~\ref{fig:code}, we show the Python implementation of the mapping function defined in Equation~\ref{eq:mapping}. This operation is $\mathcal{O}(n)$, where $n$ is the number of rows in the DataFrame, and it is scalable as it is applied independently to each row.
		\item Notice that the DataFrame with only the ``case'' and ``activity'' columns represents the activity-log according to our definition in Sec.~\ref{sec:meth}. The construction of the DFG requires a single iteration through the activity-log. Therefore, this operation is also $\mathcal{O}(n)$ and can be scaled~\cite{evermann_scalable_2016}.
		\item The computation of I/O statistics requires a single pass over the DataFrame followed by a grouping and aggregation based on activity values. Therefore, the complexity is $\mathcal{O}(mn)$, where $m$ is the number of unique activities $|\mathcal{A}_f|$ (i.e., the number of nodes in the DFG). For all intents and purposes, the mapping function should be defined such that $m$ is small; otherwise, the visual analysis of the DFG would be tedious.
		\item The DFG along with the statistics are rendered. The rendered DFG is styled by applying one of the coloring methods described in Sec.~\ref{sec:meth_coloring}. The complexity of the rendering is $\mathcal{O}(m^2)$ in the worst case; i.e., when every node has an edge to every other node. Note that partition-based coloring (Step 5b in Figure~\ref{fig:code}) requires at most one additional pass over the activity-log to construct both the DFGs, i.e., $\mathcal{O}(n)$, to compute both \textit{green\_dfg} and \textit{red\_dfg}.  
	\end{enumerate}
	
	Our implementation is available through the Python library \textit{st\_inspector}. The source code is available in Zenodo~\cite{sankaran_strace_2024}. This implementation is used for the experiments.
	
	
	\begin{figure*}[t!]
		\centering
		\begin{subfigure}[b]{1\textwidth}
			\centering
			\includegraphics[width=0.9\linewidth]{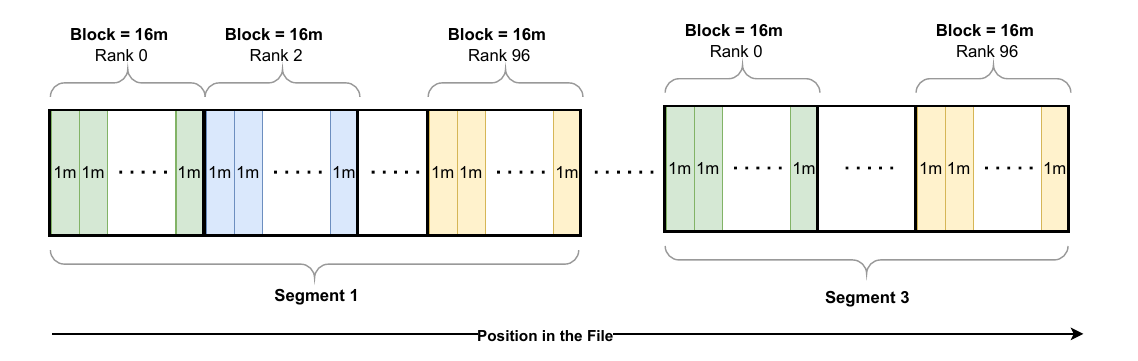}
			\caption{The format of the IOR file.}
			\label{fig:ior_file}
		\end{subfigure}
		\begin{subfigure}[b]{1\textwidth}
			\centering
			\includegraphics[width=0.6\linewidth]{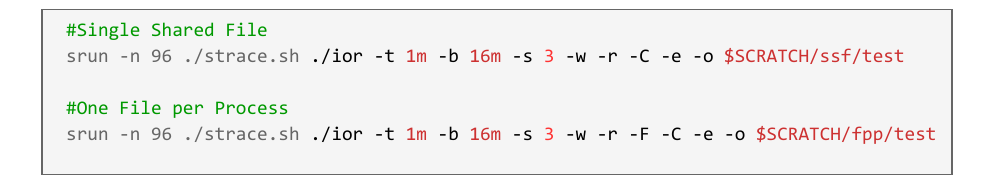}
			\caption{IOR commands to simulate SSF and FPP scenarios.}
			\label{fig:ior_cmd}
		\end{subfigure}
		
		\caption{The IOR Experiment.}
		\label{fig:eg_ior}
	\end{figure*}
	
	\textbf{The HPC Environment:} The experiments are conducted on the compute nodes of the JUWELS cluster~\cite{alvarez_juwels_2021} at the J\"ulich Supercomputing Centre. We use the nodes of JUWELS that have 2x24 cores Intel Xeon Platinum 8168 CPUs with 96 GB DDR4 memory. JUWELS is connected via Connext-X4 Infiniband/Ethernet adaptor to the $6^{th}$ generation JUST storage infrastructure~\cite{graf_just_2021}, which is a GPFS based file system.
	
	%

	\subsection{Single Shared File vs File Per Process}

	In applications where multiple processes are simultaneously involved in I/O operations, we compare the following two scenarios: (1) Single Shared File (\textbf{SSF}): all processes read from or write to a single shared file, and (2) File Per Process (\textbf{FPP}): each process accesses its own individual file. Generally, allowing each process to work on its own file eliminates the contention issues arising from file locking, which is otherwise common in the shared file scenario. However, creating a file for each process at scale leads to metadata overhead, which hits performance especially when data needs to be frequently gathered across processes. Therefore, users need to gauge the trade-offs between the two approaches and quantify, for a given application, whether contention leads to significantly increased execution times compared to the case where each process operates on its own file.  In this experiment, we apply our methodology to verify the possibility of identifying these contention issues and visualizing the differences in file access activities between the two scenarios. 
	\begin{figure*}[t!]
		\centering
		\begin{subfigure}[b]{1\textwidth}
			\centering
			\includegraphics[width=1\linewidth]{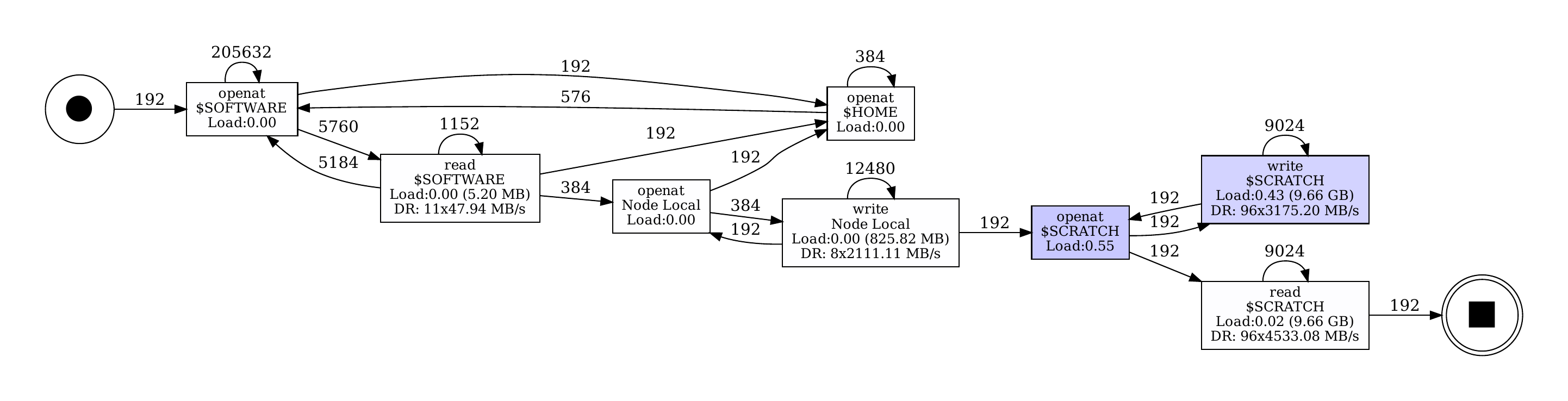}
			\caption{ DFG synthesis applied to all the events.}
			\label{fig:ssf_fpp_dfg0}
		\end{subfigure}
		\begin{subfigure}[b]{1\textwidth}
			\centering
			\includegraphics[width=0.48\linewidth]{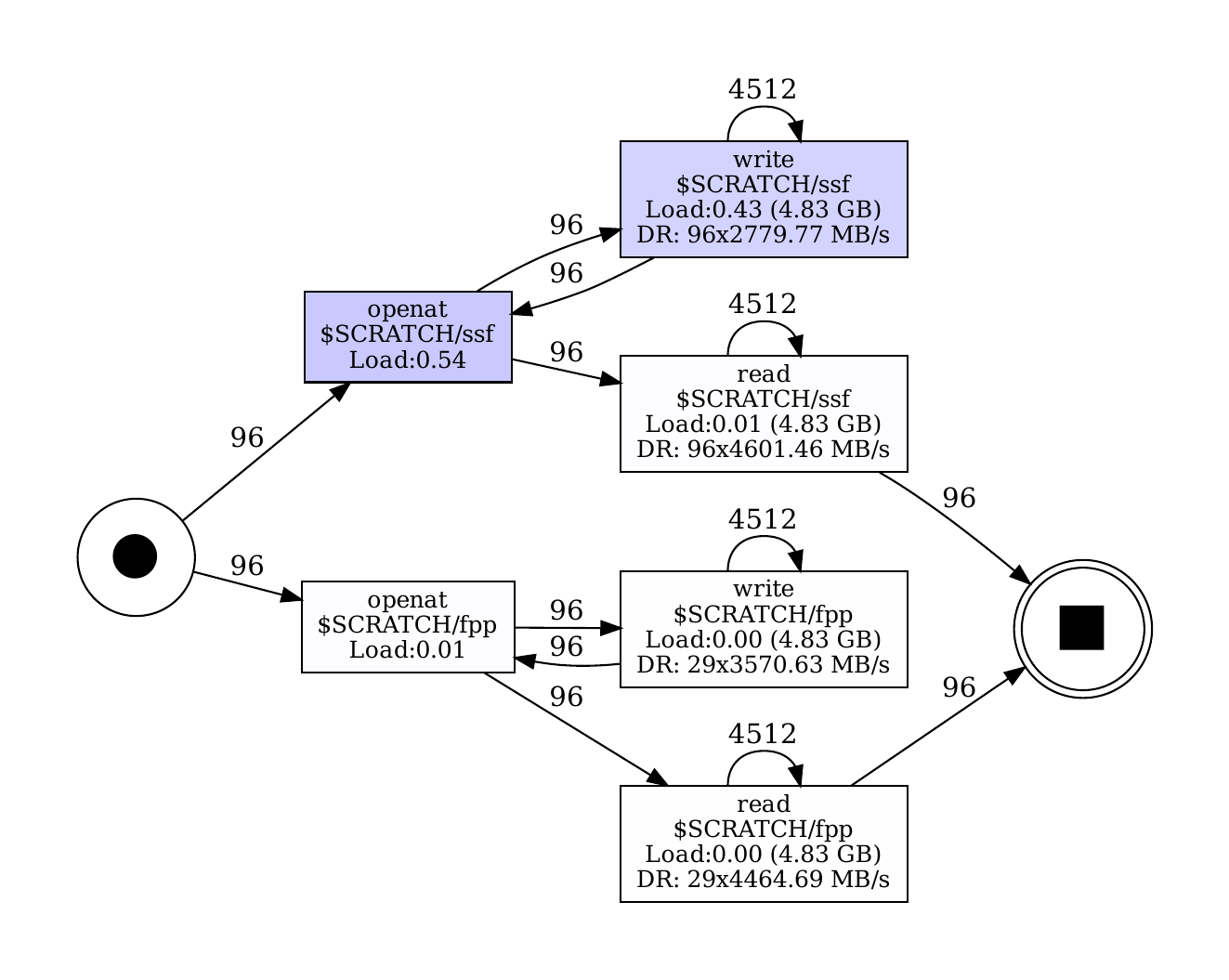}
			\caption{DFG synthesis applied only to events that access the \texttt{\$SCRATCH} directory.}
			\label{fig:ssf_fpp_dfg}
		\end{subfigure}
		
		\caption{DFG synthesis of the events from both SSF and FFP modes of IOR runs.}
		\label{fig:exp_dfg}
	\end{figure*}
	
	We use the IOR benchmark suite to write and read in parallel from 96 MPI processes or rank spanning across 2 nodes of the JUWELS cluster (i.e., 96 cores in total and we execute one MPI rank per core). Each rank first writes 3 segments of data, with each segment consisting of a 16 MB block, and each block divided into 16 write operations of 1 MB each. Subsequently, each rank reads the data written by a process from the neighboring node (this is done to avoid reading the data stored in the DRAM). The format of the file is shown in Figure~\ref{fig:ior_file}, and the IOR options for the SSF and FPP experiments are shown in Figure~\ref{fig:ior_cmd}. The number of segments, block size, and size of each operation are specified with the options \texttt{-s}, \texttt{-b}, and \texttt{-t}, respectively. The option \texttt{-C} forces the MPI ranks to read the data written by the neighboring node, and the option \texttt{-e} ensures that the write to the file system is complete before starting the subsequent read operation.  By default, IOR runs in the SSF mode, and switches to FPP when the option \texttt{-F} is specified. We run IOR in both modes; for SSF, the files are accessed from the folder \texttt{\$SCRTACH/ssf}, and for FPP, the files are accessed from the folder \texttt{\$SCRTACH/fpp}. The access path is specified using the \texttt{-o} option. We aim to identify the differences in file contentions occurring in these two folders.
	\begin{figure*}[t!]
		\centering
		\includegraphics[width=1\linewidth]{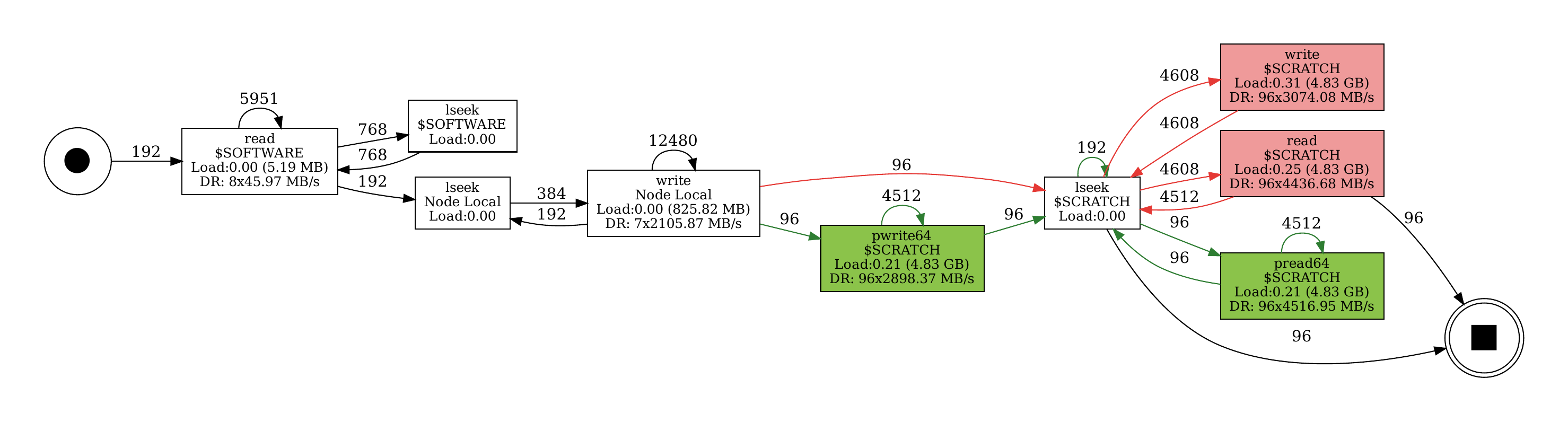}
		\caption{DFG synthesis of events from the MPI-IO experiment.}
		\label{fig:posix_mpiio_dfg}
	\end{figure*}
	
	We record the events related to variants of \texttt{read}, \texttt{write} and \texttt{openat} system calls and prepare the HDF5 event-log. The event-log $\mathcal{C}_X$ consists of 192 cases (each stored as a table in the HDF5 file); 96 from the SSF run and 96 from the FPP run. We retrieve all the events without applying any file path filters.
	The map of events to activity values ($\bar{f}$) is similar to the one defined in Equation~\ref{eq:mapping}, but abstracts the file paths based on site-specific variable. 
	
	After applying the mapping $\bar{f}$, the activity-log $L_{\bar{f}}(\mathcal{C}_X)$ is determined, and the DFG $G[L_{\bar{f}}(\mathcal{C}_X)]$ is constructed, as shown in Figure~\ref{fig:ssf_fpp_dfg0}. The I/O statistics described in Section~\ref{sec:meth_stats} are computed and the nodes of the DFG are colored based on the corresponding relative duration statistic ($rd_{\bar{f}}$) computed according to Equation~\ref{eq:rd}; higher the value of $rd_{\bar{f}}$ , darker the shade of blue. It can be seen that \texttt{openat} and \texttt{write} operations under the \texttt{\$SCRATCH} directory have a relatively high load.

	Now, knowing that there is a relatively high load under the \texttt{\$SCRATCH} directory, we filter the event-log to retrieve only those events that access the \texttt{\$SCRATCH} directory. We re-apply the mapping and construct the DFG, which is shown in Figure~\ref{fig:ssf_fpp_dfg}. It can be seen that the \texttt{openat} and \texttt{write} activities on files under \texttt{\$SCRATCH/ssf} (which is the access path for the IOR run in SSF mode)  have a significantly higher relative duration than those corresponding to the files under \texttt{\$SCRTACH/fpp} (access path for the IOR run in FPP mode). This quantifies the contention issue due to file locking in the SSF scenario in terms of execution times.

	\subsection{With vs Without MPI-IO Interface}

	The MPI-IO provides standard interfaces for parallel I/O access~\cite{message_passing_interface_forum_mpi-2_1997}. 
	It has been noted that performance gains through this interface are not guaranteed unless it is correctly configured by identifying the MPI-IO defined file access patterns within the program~\cite{thakur_case_1998}. Therefore, tools that assist users in comparing the performance impacts of different interface configurations can be beneficial. In this experiment, we run the IOR benchmark both with and without the MPI-IO interface. By default, IOR does not use the MPI-IO interface. Adding the option  \texttt{-a mpiio} would do a naive replacement of standard file operations with the MPI-IO counterpart, without the use of advanced MPI-IO configurations. We run IOR in SSF mode as before, this time with and without the \texttt{-a mpiio} option, and compare the file access contention between the runs.

	In addition to variants of \texttt{read}, \texttt{write}, and \texttt{openat}, we also record the events related to \texttt{lseek}, and prepare the event-log $\mathcal{C}_Y$. We retrieve all the events, apply the mapping $\bar{f}$,  construct the DFG  $G[L_{\bar{f}}(\mathcal{C}_Y)]$, and compute the I/O statistics. Unlike the previous run, this time the two runs do not use distinct file access paths. Therefore, to compare the two runs, we apply the partition-based coloring described in Sec.~\ref{sec:meth_coloring}. To this end, the event-log $\mathcal{C}_Y$ is partitioned into two  mutually exclusive event logs, $\mathcal{G}_Y$ and $\mathcal{R}_Y$. The event-log $\mathcal{G}_Y$ constitutes cases that were run \textit{with} the MPI-IO interface, and  $\mathcal{R}_Y$ constitutes cases that were run \textit{without} the MPI-IO interface.  We color the nodes of $G[L_{\bar{f}}(\mathcal{C}_Y)]$ according to the steps defined in  partition-based coloring (Sec.~\ref{sec:meth_coloring}). The resulting DFG is shown in Figure~\ref{fig:posix_mpiio_dfg}; the green nodes and edges are those that occur exclusively in the run with MPI-IO interface, and red nodes and edges are those that occur exclusively in the run without the MPI-IO interface. All other nodes and edges occur in both the runs (we skip the rendering of \texttt{openat} calls in Figure~\ref{fig:posix_mpiio_dfg} as it does not highlight useful differences).

	It can be seen that MPI-IO utilizes the system calls \texttt{pread64} and \texttt{pwrite64}  instead of the standard \texttt{read} and \texttt{write}. Standard \texttt{read} or \texttt{write} calls start from the offset left by the previous access in an opened file, which means other processes must call \texttt{lseek} to reset the offset to the correct position before performing their own file operations. The \texttt{pread64} and \texttt{pwrite64} system calls combine file access and seek operations into a single command. This eliminates the need for an explicit \texttt{lseek} call before reading or writing, thereby reducing the number of system calls issued from the user environment.  
	Therefore, one could observe that the number of \texttt{lseek} calls preceding file accesses is significantly lower in the run that use MPI-IO interface compared to the run without MPI-IO. We observe that the reduction in the number of system calls resulted in a relatively reduced load in terms of overall duration.
	
	\textbf{Availability of Data and Materials: } The experimental data and the results that support the findings of this study are available in Zenodo with the identifier \href{https://doi.org/10.5281/zenodo.13325645}{https://doi.org/10.5281/zenodo.13325645}.

	\section{Conclusion}
	\label{sec:con}

	In this paper, we presented a dependency-graph-based methodology for comparative analyses of arbitrary user programs in terms of I/O requests made to the operating system. To this end, we considered the I/O operations from the traces of system calls of one or more programs as input. We presented the methodology to transform the input data into a Directly-Follows-Graph consisting of nodes that represent I/O operations occurring at specific file paths, and the edges representing the directly-follows relation between those I/O operations. Based on the Directly-Follows-Graph, we described the technique to compare and contrast the patterns of I/O operations between multiple programs. We tested our methodology by applying it to the IOR benchmark and validating the similarities and differences in the patterns of I/O requests made to the operating system when IOR was run with different options for file output and software interface.
	
	Our methodology can be used in tools that aim to diagnose and compare programs based on system resource contentions to identify I/O performance bottlenecks. Such tools are particularly valuable for supporting users of HPC systems and ensuring optimal resource usage. In future work, we plan to apply our technique to typical HPC workloads and document the findings.
	
	\bibliographystyle{IEEEtran}
	\bibliography{references}

\end{document}